\documentclass[twocolumn,prl,footinbib]{revtex4}

\usepackage{graphicx}
\usepackage{dcolumn}
\usepackage{bm}
\usepackage{hyperref}

\begin{document}

\title{High fidelity quantum gates via analytically solvable pulses}

\author{Sophia E. Economou}

\affiliation{Naval Research Laboratory, Washington, DC 20375, USA}
\date{\today}

\begin{abstract}
It is shown that a family of analytically solvable pulses can be
used to obtain high fidelity quantum phase gates with surprising
robustness against imperfections in the system or pulse parameters.
Phase gates are important because they can implement the necessary
operations for universal quantum computing. They are particularly suited for
systems such as self-assembled quantum dots, trapped ions, and
defects in solids, as these are typically manipulated by the
transient excitation of a state outside the qubit subspace.
\end{abstract}

\maketitle

The physical implementation of quantum computation requires high
quality coherent gates. Single qubit rotations combined with the
conditional C-Z phase gate form a universal set of quantum logic
gates \cite{nielsenchuang}. Time dependent controls, such as lasers,
are used in order to implement these prescribed unitary evolutions
of the qubits.

In many physical systems, such as self-assembled quantum dots
\cite{press,spinrot,nrlwork,godden} and quantum wells
\cite{hwang}, trapped ions \cite{campbell} and atoms
\cite{atomsreview}, defects in solids \cite{buckley}, and in some
cases superconducting qubits \cite{dicarlo}, auxiliary states
outside the Hilbert space of the qubit are used in order to
implement these gates. One advantage of using such states is that
their energy splitting from the qubit states is typically orders of
magnitude larger than the qubit splitting itself and thus fast
operations can be achieved, since the time scales as the inverse of
the energy. For unitary evolution within the qubit space, these
excited states should be only transiently excited, and fast
operations are key in order to avoid incoherent decay back to the
qubit states.

The most familiar gate is the induction of a minus sign
in front of one of the qubit states via cyclic resonant excitation
of the excited state. As shown in Fig. 1, the two qubit states
$|n\rangle$ and $|\bar n\rangle$ (which can be thought of as spin
eigenstates along the $\mathbf{n}$ direction, or as a subset of
two-qubit states) are respectively coupled and uncoupled from the
excited state by a pulse. A path that excites the population in
state $|n\rangle$ resonantly to $|E\rangle$ and returns it through a
cyclic evolution will induce a minus sign to state $|n\rangle$. This
is a familiar property of quantum systems that differentiates (pseudo)spin from spatial rotations in
euclidian space, where a full $2\pi$ rotation returns the system to
its starting point.

Indeed, this evolution has been used in the demonstration of quantum
gates of a variety of systems, including semiconductor
nanostructures \cite{spinrot,hwang,ekim,gershoni}, trapped ions
\cite{cnot_ions} and fullerene molecules \cite{morton}. This is done
straightforwardly by a resonant pulse of any temporal profile. In
contrast, the implementation of other phase gates is non trivial, since the majority of pulses will
leave the system partially in the leaky excited state $|E\rangle$. Perturbative methods such as adiabatic elimination \cite{SW} of state $|E\rangle$
partially address this, but they have the inherent drawback of a need for long pulses. Exact
analytically solvable dynamics are therefore highly desirable, since
they can guarantee that the probability of the
population remaining in the excited state is zero after the passage
of the pulse. Moreover, they provide a reliable recipe for tuning parameters
to achieve the target evolution. In that spirit, the well-known
hyperbolic secant (sech) pulse \cite{rosenzener} was proposed
\cite{economouprbprl} and later used experimentally for the
demonstration of electron \cite{spinrot} and exciton \cite{gershoni} spin
rotations in quantum dots.

While the sech pulses considered in
\cite{economouprbprl,spinrot,gershoni} have been successful for the
experimental demonstration of spin gates, they do not yield the near
perfect fidelities needed for quantum computing. The main
shortcomings of these pulses are the (near) resonant excitation of
the lossy excited state required to achieve large phases and the
sensitivity of the induced phase to small variations in the
detuning. As a result, an experimental error in the laser frequency
induces errors in phase, while in an ensemble of
systems with unequal energy splittings this can lead to dephasing in
the net signal \cite{spinrot}.

In this work, these issues are addressed by use of a family of
pulses that generally feature asymmetric profiles and frequency
modulation (`chirp'). Focusing on cyclic evolutions by two different
pulse strengths, I show that impressive
fidelities are obtained by the stronger pulses, which
furthermore demonstrate robustness against errors in the parameters.
The chirped pulses generally allow for higher fidelities as compared
to their unchirped counterparts, an effect reminiscent of the robust
population transfer to an excited state using chirped lasers, which
has been recently used to that end in quantum dot systems
\cite{ywu_chirp,simon}. In the context of the latter experiments,
the present work is particularly timely and compatible with state of
the art capabilities.
\begin{figure}[htp]
\includegraphics[width=1.8 cm,clip=true, bb=10.6cm 5.5cm 14.5cm 11.1cm]{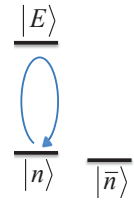}
\caption{(Color online) Qubit states ($|n\rangle, |\bar n \rangle$)
and the auxiliary excited state $|E\rangle$. States $|n\rangle$ and
$|E\rangle$ have energy separation $\omega_{En}$ and are coupled by
a field of frequency $\omega$. } \label{levels}
\end{figure}

The hyperbolic secant (sech) pulse shape was first discovered by
Rosen and Zener (RZ) as a nontrivial driving term that yields an
analytically solvable Schrodinger equation of a two-level system. In
the 70s the sech pulse gained intense interest in the context of
self induced transparency, i.e., as the pulse that does not lose its
shape as it propagates through a nonresonant medium \cite{sit}. In
the 80s a number of papers introduced pulses that also result in
analytically solvable dynamics and which are generalizations of the
sech, generally containing asymmetry and/or chirping
\cite{bambiniberman,zakrzewski}. It was found \cite{bambiniberman}
that asymmetric pulses do not return the population to the ground
state, while the same is true for ones with only chirping. However,
the combination of chirp and asymmetry allows for cyclic evolution
when the chirp and asymmetry parameters are related in a certain way
\cite{zakrzewski}.

To solve the time-dependent Schr\"{o}dinger equation for a coupling
with a hyperbolic secant temporal shape $f(t)=\Omega
\text{sech}(\sigma t) e^{i\omega t}+cc$, RZ define a new variable
$z(t)=1/2(1+\tanh(\sigma t))$. In terms of $z$, they show that the
second order differential equation for the probability amplitude of
state $|n\rangle$ is the hypergeometric equation, for which the
solutions are known. From the solution, one can see that when
$\Omega/\sigma=$integer, the evolution is cyclic, independently of
the detuning $\Delta=\omega_{En}-\omega$ from the transition
$|n\rangle\rightarrow|E\rangle$.

In the more general case of temporally asymmetric and frequency
modulated pulses with envelope $f(t)=2\sqrt{z(1-z)}/(\lambda z+1)$
and oscillatory term $e^{i(\omega t+g(t))}$, where
$\dot{g}=\beta\frac{(2+\lambda)z-1}{\lambda z+1}$, a similar
transformation $t=\sigma^{-1}/2 \ln[z/(1-z)^{1+\lambda}]$ with
$\lambda>-1$ puts the equation in a Hypergeometric form. Therefore,
the dynamics are analytically solvable, and what changes relative to
the RZ solution are the parameters appearing in the Hypergeometric
functions, which are now functions of not only the pulse strength
and frequency, but also the asymmetry and the chirp. In this general
case, the probability of return does depend on the detuning as well
in the following way: when the chirp parameter $\beta$ is related to
the asymmetry via $\beta=-\lambda\Delta/(2+\lambda)$, the effective
pulse area is the same as that of the RZ sech pulse, so that for
$\Omega/\sigma=$ integer, the induced evolution is cyclic. I focus
on cyclic evolution and specifically on $\Omega=\sigma$ and
$\Omega=2\sigma$, which will be referred to as 2$\pi$ and 4$\pi$
pulses respectively. A 2$\pi$ (4$\pi$) pulse cycles the polarization vector
from $|n\rangle$ toward $|E\rangle$ and back to $|n\rangle$ once (twice). Taking $\lambda=0$ recovers the symmetric,
unchirped sech pulses.

A cyclic evolution in general induces a phase to the state, i.e.,
the probability amplitude acquires a phase factor. Using the
analytically solvable dynamics outlined above, I obtain the
following expressions for the phases for 2$\pi$ and $4\pi$ pulses
respectively
\begin{eqnarray}
\phi(2\pi) &=& 2\arctan\left[\frac{2+\lambda}{2(1+\lambda)}\frac{\sigma}{\Delta}\right]\label{f2p}
\\
\phi(4\pi) &=& 2\arctan\left[\frac{8(1+\lambda)(2+\lambda)\Delta/\sigma}{4(\Delta/\sigma)^2(1+\lambda)^2-3(2+\lambda)^2}\right].\label{f4p}
\end{eqnarray}

\begin{figure}[htp]
\includegraphics[width=6.8cm]{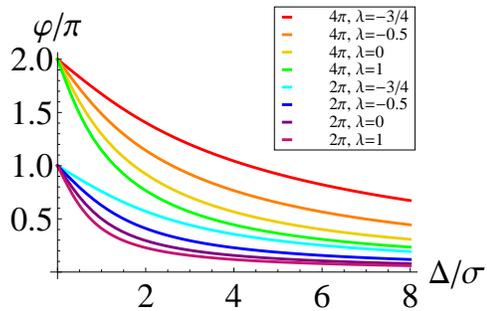}
\caption{(Color online) Phase (modulo 2$\pi$) as function of the
detuning in units of pulse bandwidth for 4$\pi$ and 2$\pi$ pulses
for various asymmetry parameters as shown in the legend. }
\label{angle}
\end{figure}

In general, the pulses are defined through their pulse shape
asymmetry and their chirp, but since the two are related for
transitionless dynamics henceforth only the strength and asymmetry
will be used to refer to each pulse. To examine the qualitative
behavior of these pulses, I focus on three non trivial values for
the asymmetry parameter, $\lambda=-3/4,-0.5,1$ for which $z(t)$ can
be inverted. The phase $\phi$ from Eqs. (\ref{f2p}) and (\ref{f4p})
is plotted against the detuning (in units of bandwidth) in Fig.
\ref{angle}, where one may notice that for a target phase $\phi$,
the phase is a slower varying function for the $4\pi$ set of pulses.
Within that set, the curves with negative $\lambda$'s are even more
slowly varying. In the limit $\lambda\rightarrow -1$ we get a
horizontal straight line, which means that the phase is always $-1$
and $1$ for 2 and 4 $\pi$ pulses respectively. This is somewhat
reminiscent of adiabatic rapid passage, where chirped pulses invert
population robustly in the presence of variations in detuning. Moreover, 4$\pi$ pulses
have larger detunings for the same phase, resulting in less real
excitation of the leaky excited state.

\begin{figure}[htp]
\includegraphics[width=7cm]{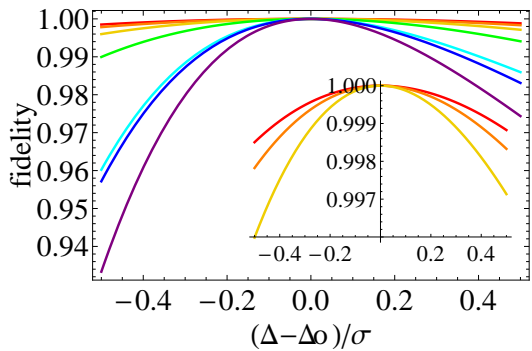}
\caption{(Color online) Fidelity of $\pi/2$ rotation as function of
deviation from the ideal detuning $\Delta_o$ for, from top to
bottom, 4$\pi$ $\lambda=-3/4$, 4$\pi$ $\lambda=-0.5$, 4$\pi$
$\lambda=0$, 4$\pi$ $\lambda=1$, 2$\pi$ $\lambda=-3/4$, 2$\pi$
$\lambda=-0.5$, and 2$\pi$ $\lambda=0$. The pulse 2$\pi$ $\lambda=1$
is not included here as it performs worse than the 2$\pi$ sech, as
expected from the plots in Fig. \ref{angle}. The inset shows the
fidelities for only the three best performing $4\pi$ pulses in a
scale where they can be distinguished.} \label{detuning}
\end{figure}

In the present case, we expect the flatter pulses to be more robust
against errors in the detuning. This statement is checked by
calculating the fidelity of the phase gate implemented with each
pulse as a function of the deviation from the ideal detuning. The
fidelity is a measure of how close our evolution $U$ is to the
target evolution $U_t$, and is defined as
$\overline{|\langle\psi|U^\dagger U_t|\psi\rangle|^2}$, where the
average is taken over all possible initial states $|\psi\rangle$.
Indeed, calculations of the fidelity  verify this expectation and
quantify the performance of these gates, as can be seen from Fig.
\ref{detuning}, where $\Delta_o$ is the ideal detuning for the
target transition and is given in terms of the target phase and the
pulse parameters $\sigma,\lambda$ by inverting Eqs. \ref{f2p} and
\ref{f4p}. Even though in the case of chirped asymmetric pulses the
deviation from the ideal detuning affects the probability of
complete return to the ground state, these pulses are still more
robust than their symmetric, unchirped counterparts. The implication
of this result is a highly desirable robustness against errors in
the pulse frequency or uncertainty of system parameters. Furthermore, this
insensitivity in the detuning in Fig. \ref{detuning} tells us that we can
implement high fidelity quantum gates in an inhomogeneous ensemble
of systems by use of a single pulse. This is particularly important
for human-made systems such as quantum dots, but it can also impact
naturally occurring systems that interact with slightly different
environments.

\begin{figure}[htp]
  \includegraphics[height=3cm, width=.48\columnwidth]{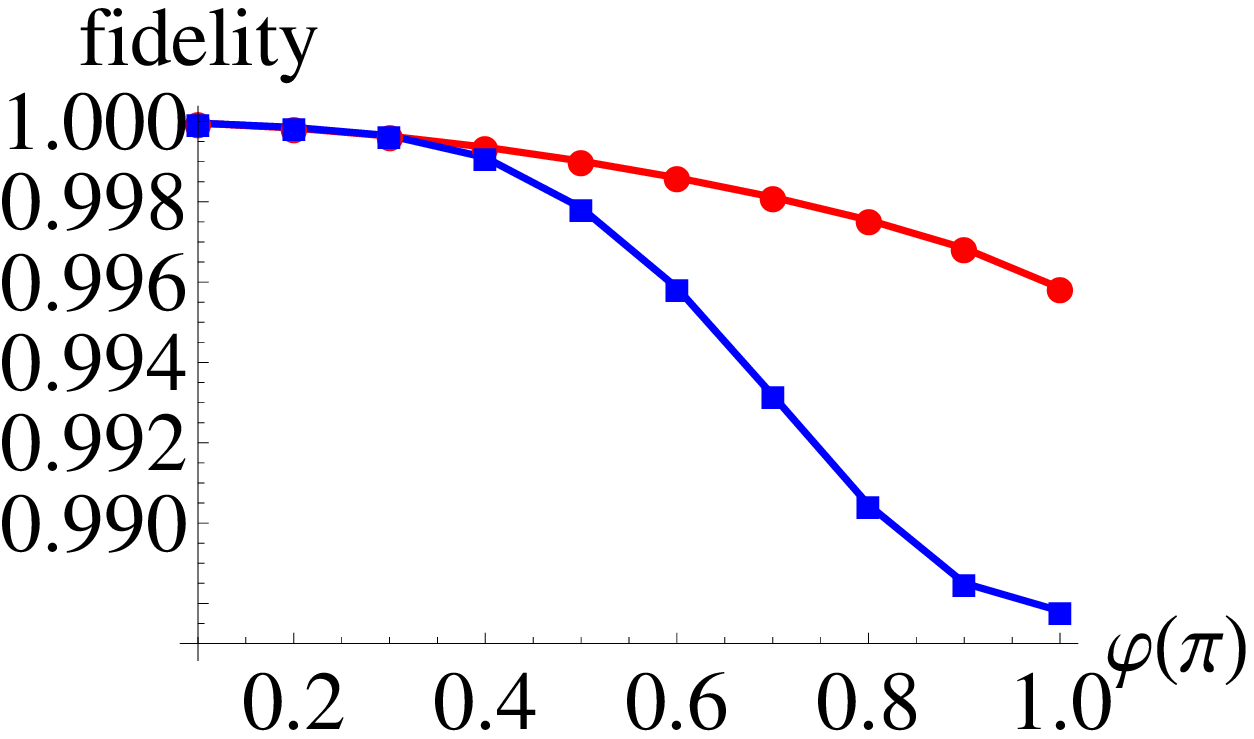}
  \includegraphics[height=3cm, width=.48\columnwidth]{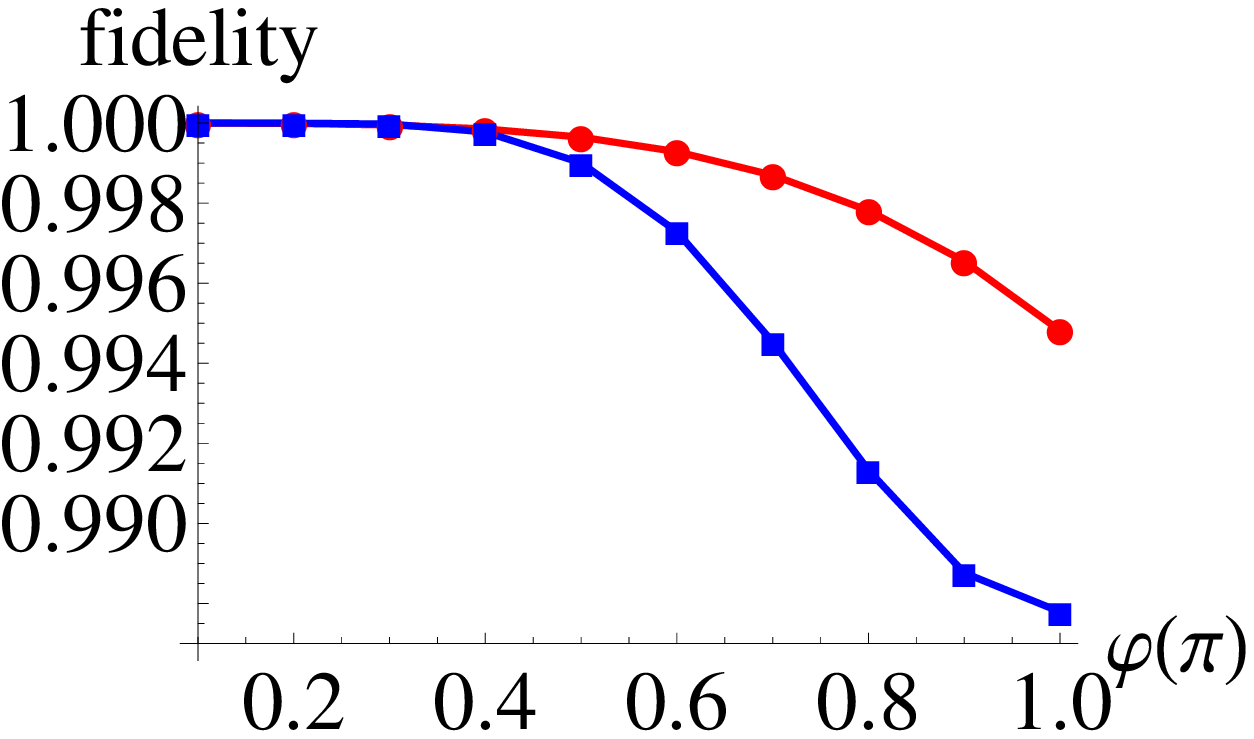}
  \caption{(Color online) Left panel: Fidelity as function of the phase for 4$\pi$ (red, circles)
and 2$\pi$ (blue, squares) pulses. The dipole of the transition is
5\% higher than the ideal value. The $4\pi$ pulse is more robust to
errors in the dipole. Right panel: Fidelity as function of the phase
for 4$\pi$ (red, circles) and 2$\pi$ (blue, squares) pulses. The
Zeeman splitting between the energy eigenstates of the qubit
subspace is 10 times smaller than the bandwidth.}\label{figjnt}
\end{figure}

Another possible source of error can be the strength of the dipole
of the transition. The various $4\pi$ pulses with different values
of $\lambda$ perform almost identically, while there is no
significant difference within the $2\pi$ pulse subset either.
However, the $4\pi$ pulses are overall superior to the $2\pi$ ones.
This can be seen in the left panel of Fig. \ref{figjnt}, where the
fidelity is shown as a function of the phase for a $4\pi$ and for a
$2\pi$ sech pulse with the dipole of the transition set to 1.05
times its target value. The superiority of the $4\pi$ pulse can be
traced back to the larger detuning required for the same phase.
Since the detuning is large, from a qualitative effective Rabi
frequency argument, the relative importance of the coupling strength
compared with the detuning is small for 4$\pi$ pulses. However, for
$2\pi$ pulses, where the control is almost resonant (and exactly
resonant for a $\pi$ rotation), the deviation from the ideal
coupling strength will have a greater effect.

Now let's consider the case where state $|n\rangle$ is not an energy
eigenstate, but instead a linear superposition of the two energy
eigenstates of the qubit subspace. This is a common situation in
quantum dot spin qubits, where there is an external magnetic field
in the plane of the dot, while the laser propagates perpendicular to
that along the growth axis and is circularly polarized. In that
case, the light couples to the `up' spin state along the growth
direction, which is not an energy eigenstate. The common assumption
and experimental practice is to take the pulse duration to be much
faster than the spin precession, so that the picture of Fig.
\ref{levels} is valid (Otherwise the coupling between states
$|n\rangle$ and $|\bar n\rangle$ should be taken into account, and
the problem is no longer analytically solvable). It is therefore
natural to ask what is the fidelity of the present pulses when the
Zeeman splitting is not negligible compared to the pulse bandwidth.
Again, the family of $4\pi$ pulses is found to perform better than
their $2\pi$ counterparts \cite{pei}, as can be seen in the right
panel of Fig. \ref{figjnt}. Since for the $4\pi$ pulses both the
detuning and the pulse strength are larger than that of the $2\pi$
pulses, the relative importance of a larger Zeeman term is smaller
in the former case, and hence the better performance.

Finally an important situation to examine is the case where there
are other excited states that the laser couples to. In principle,
these can be avoided by making the pulses temporally very long,
i.e., very narrowband. This is generally not practical however,
since long pulses excite the lossy auxiliary state $|E\rangle$ for
longer times, increasing its chance to decay. Thus the compromise is
to pick pulses that are narrowband compared to the energy difference
between the target and unwanted transition, and broadband compared
to the linewidth of the excited state. For such slower pulses, the
fidelity is lowered predominately due to incoherent dynamics. This
can be seen by calculating the purity of the qubit state after the
passage of the pulse.

\begin{figure}[htp]
  \includegraphics[height=3cm, width=.48\columnwidth]{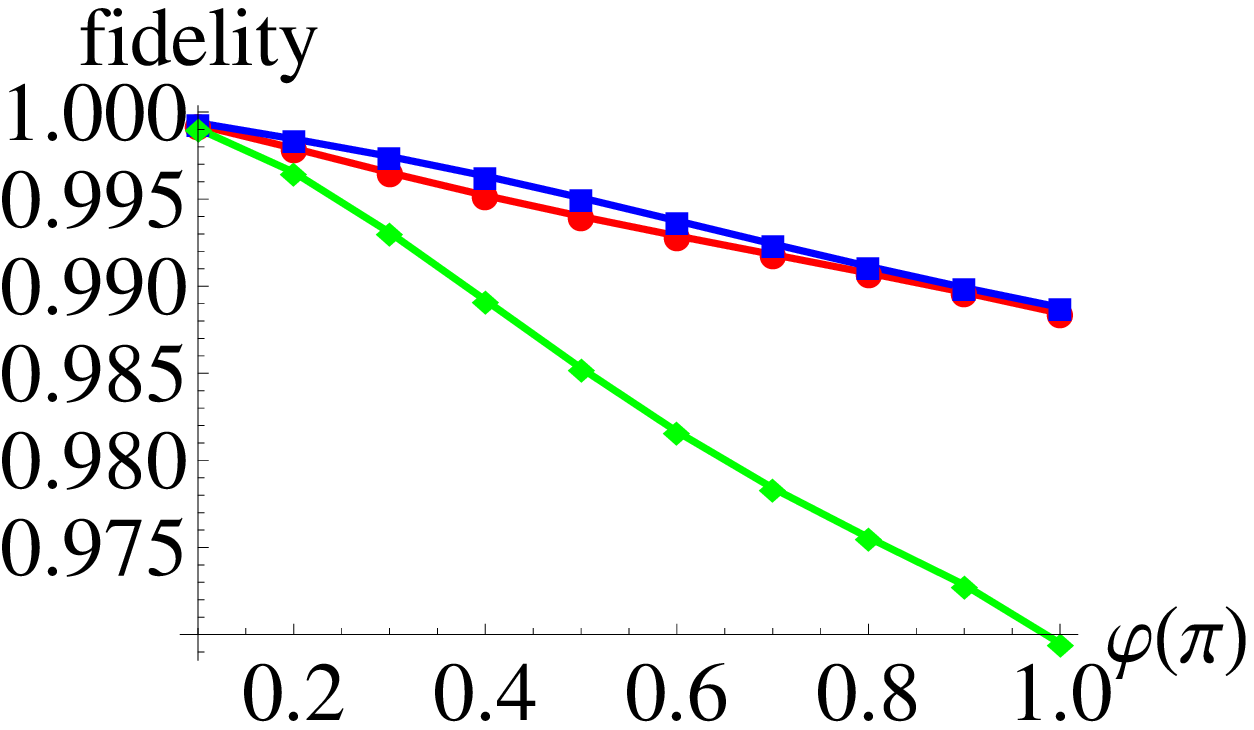}
  \includegraphics[height=3cm, width=.48\columnwidth]{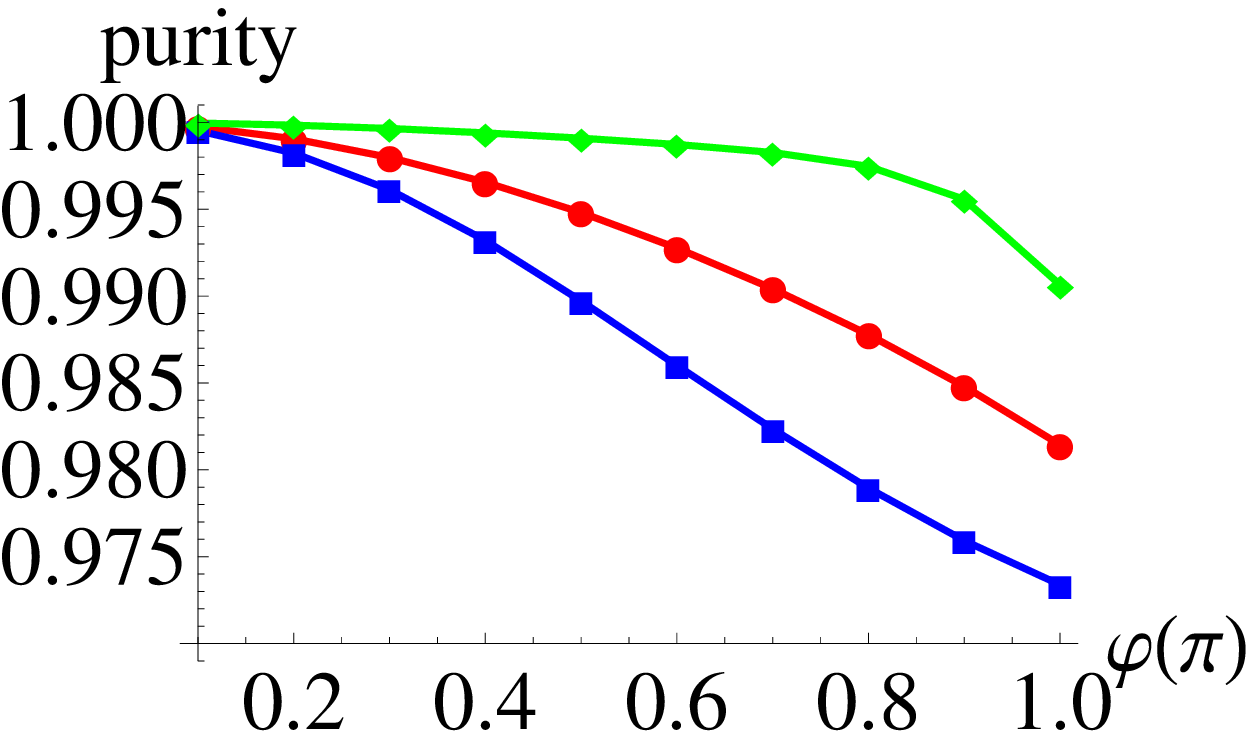}
  \caption{(Color online) Gate performance in the presence of additional
 excited state $|E'\rangle$ and decay. The transition
$|n\rangle\rightarrow|E'\rangle$ is taken to be 0.2 meV higher in
frequency from $|n\rangle\rightarrow|E\rangle$ and to have half the
coupling strength (qualitatively similar results are obtained for
both higher and lower coupling strengths). The decay rates are set
to 0.8$\mu$eV. Left panel: Fidelity as function of the phase for
4$\pi$ pulse with $\sigma=0.03$meV (red, circles) and 2$\pi$ pulse
with $\sigma=0.06$meV (blue, squares). The green (diamonds) curve
corresponds to the fidelity of a fast 4$\pi$ sech pulse with
$\sigma=0.2$meV. Right panel: Purity as function of the phase for
4$\pi$ (red, circles) and 2$\pi$ (blue, squares) pulses with the
maximal fidelity, see left panel. The green (diamonds) curve is the
maximal purity for a 4$\pi$ sech pulse. Note that the parameters
used here are taken from quantum dot spin qubits, but are consistent
with atomic/ionic qubits since the energy scales in those systems
are all about an order of magnitude smaller.}\label{figjnt2}
\end{figure}

The purity is a measure of the incoherent dynamics and can be
defined as $Tr(\rho^2)$, where $\rho$ is the density matrix of the
qubit. Clearly, for a system in a pure state, which can be described
by a wave function, the purity is 1. It is instructive to look at
the purity and fidelity of 2$\pi$ versus $4\pi$ pulses. For this
calculation, an additional excited state $|E'\rangle$ is included in
the Hamiltonian (see caption of Fig. \ref{figjnt2}). First the
bandwidths for which the fidelity is maximized are found numerically
for each of the 2$\pi$ and $4\pi$ pulses (this happens for different
bandwidths for each case), and the corresponding purity is
calculated. The low values of the latter, as shown in the right
panel of Fig. \ref{figjnt2}, indicate that indeed most of the
fidelity loss is due to incoherent dynamics.

It is particularly interesting that while the maximal purity and
fidelity coincide at a certain value of $\sigma$ for the $2\pi$
pulse, they occur at different values of $\sigma$ for the $4\pi$
pulse. Moreover, for the latter the maximal purity is very high,
much larger than that of the $2\pi$ pulse, at a bandwidth that is
comparable to the splitting from the unwanted transition. What this
tells us is that the loss of fidelity comes almost exclusively from
unintended dynamics, i.e., that crucially the population is returned
to the qubit subspace once the pulse has passed. This surprising
result is important because it means that there exists a fast high
fidelity phase gate corresponding to this pulse. So essentially we
have unitary but unknown dynamics. This observation opens up the key
question of how to explicitly determine the actual, almost unitary
evolution corresponding to the fast, high purity pulse. The
analytically solvable dynamics presented here, in combination with
approximate methods such as split-operator techniques or other
expansions, offer a promising starting point for determining the
evolution operator when additional excited state dynamics are
involved. This pursuit would advance the design of quantum controls
for a variety of realistic systems.

In conclusion, I have shown that high fidelity phase gates can be
achieved in realistic systems by use of a family of analytically
solvable pulses that may have asymmetric temporal shapes and frequency
modulation. The superior quality of these controls, as quantified by
the fidelity, stems from their robustness against imperfections. Moreover, the relatively
simple analytical expressions derived here greatly facilitate the design of
experimental implementation of phase gates. The latter are important as
they suffice for universal quantum logic. In addition, this work opens up the possibility of
incorporating off-resonant states into the control schemes, thus
allowing for simultaneously fast and accurate control, an attractive
feature when limited coherence times are available.

This work was supported by LPS/NSA and in part by ONR. I thank Edwin Barnes
for fruitful discussions.

\end{document}